\documentclass{appolb}
\usepackage{graphicx}

 \newcommand\beq{\begin{equation}}
 
 \newcommand\eeq{\end{equation}}
 \newcommand\beqn{\begin{eqnarray}}
 \newcommand\eeqn{\end{eqnarray}}

\def\GeV{\,\mbox{GeV}}

\def\Pom{{\rm I\!P}}

\def\lsim{\mathrel{\rlap{\lower4pt\hbox{\hskip1pt$\sim$}}
    \raise1pt\hbox{$<$}}}         %less than or approx. symbol
\def\gsim{\mathrel{\rlap{\lower4pt\hbox{\hskip1pt$\sim$}}
    \raise1pt\hbox{$>$}}}         %greater than or approx. symbol

\begin{document}
\title{Production of neutrons in the vicinity of the pion pole\thanks{Presented by B.Z.K. at the 16th conference on Elastic and Diffractive scattering, EDS Blois, June 29 - July 4, 2015, Borgo, Corsica.}}

\author{
B. Z. Kopeliovich, I. K. Potashnikova, Iv\'an Schmidt
\address{Departamento de F\'{\i}sica,
and
Centro Cient\'ifico-Tecnol\'ogico de Valpara\'iso\\
Universidad T\'ecnica Federico Santa Mar\'{\i}a\\
Casilla 110-V, Valpara\'iso, Chile}
}
\maketitle
\begin{abstract}
High-energy hadronic reactions with  proton-to-neutron transitions (and vice versa) with small momentum transfer allow to study the properties of nearly on-shell pions, which cannot be accessed otherwise.
We overview the recent results for such processes in deeply inelastic scattering, single and double leading neutron production in pp collisions, including polarization effect. A special attention is paid to the absorption effects, which are found to be much stronger than what has been evaluated previously.
\end{abstract}
  
\section{Preface}

Nucleons are known to carry intensive pion clouds \cite{tony}, which can be employed to study the pion properties. In particular, leading neutron production was measured in deep-inelastic scattering (DIS) at HERA \cite{zeus,h1} aiming at extraction from data the pion structure function at small Bjorken $x$. In proton-proton collisions leading neutron production
also offers an access to the pion-proton total cross section, which has been directly measured so far with pion beams within a restricted energy range in fixed-target experiments. With modern high-energy colliders the energy range for pion-nucleon collisions can be considerably extended.
Measurements with polarized proton beams supply more detailed information about the interaction dynamic. Eventually, one can employ the unique opportunity to study pion-pion interactions in double-leading-neutron production in $pp$ collisions. These processes, data and theoretical developments, are briefly overviewed below.

\section{Leading neutrons in DIS}

\begin{figure}[htb]
\centerline{%
\includegraphics[width=5.5cm]{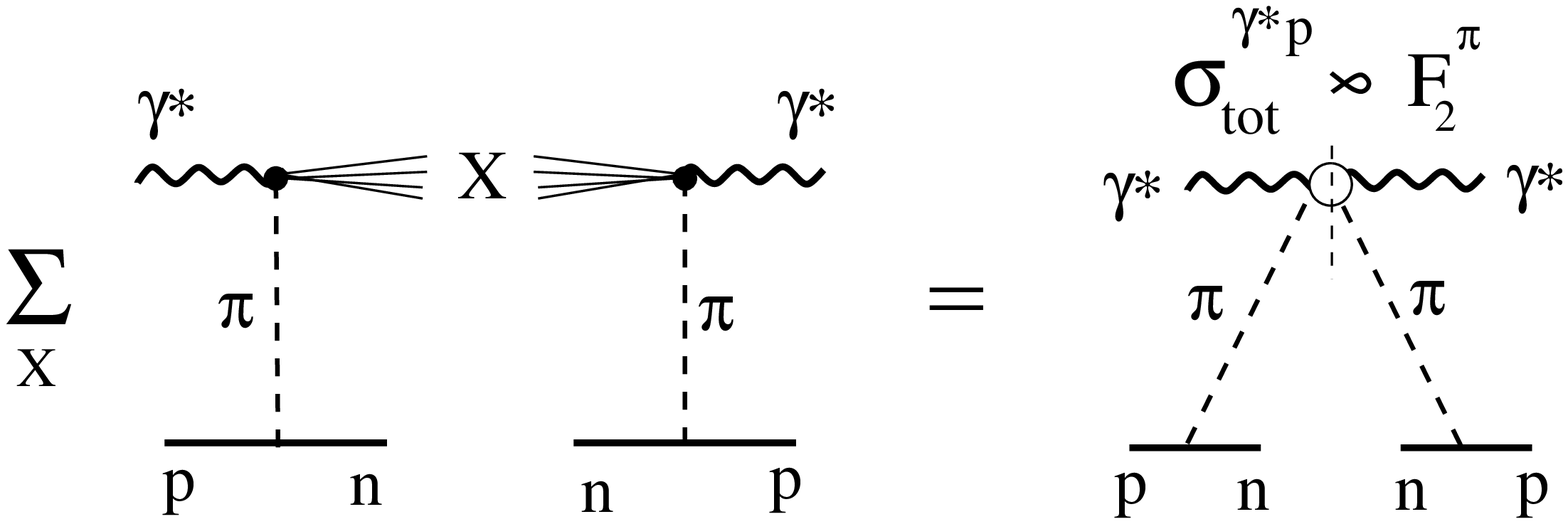}
\hspace{10mm}
\includegraphics[width=5.5cm]{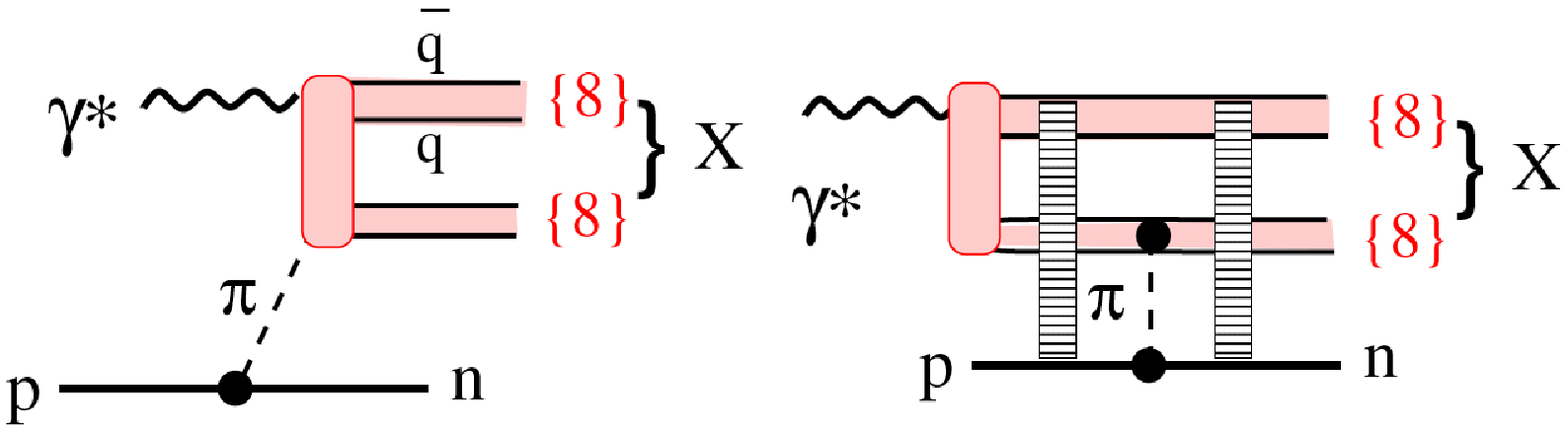}}
\caption{{\it Left:} graphical representation of the pion pole contribution to $\gamma^*p\to Xn$. {\it Right:} absorption due to interaction of the debris from the $\gamma^*\pi$ inelastic collision.}
\label{fig:pion-pole}
\end{figure}

Fig.~\ref{fig:pion-pole} (left)  illustrates how the pion structure function can be measured in the reaction $\gamma^*p\to Xn$.
%%%%%%%%%%%%%%%%%%%%
%%%%%%%%%%%%%%%%%%%%
The amplitude of this process in the Born approximation (no absorption corrections) has the form \cite{kpps-dis},
 \beq
A^B_{p\to n}(\vec q,z)=
\bar\xi_n\left[\sigma_3\, q_L+
\frac{1}{\sqrt{z}}\,
\vec\sigma\cdot\vec q_T\right]\xi_p\,
\phi^B(q_T,z)\,,
\label{100}
 \eeq
where $\vec\sigma$ are Pauli matrices;  $\xi_{p,n}$ are the proton or
neutron spinors;  $\vec q_T$ is the transverse momentum transfer;
$q_L=(1-z)m_N/\sqrt{z}$; and $z$ is the fractional light-cone momentum of the initial proton, carried by the final neutron.

 At small $1-z\ll1$ the pseudo-scalar amplitude $\phi^B(q_T,z)$ has
the triple-Regge form \cite{kpp},
 \beq
\phi^B(q_T,z)=\frac{\alpha_\pi^\prime}{8}\,
G_{\pi^+pn}(t)\,\eta_\pi(t)\,
(1-z)^{-\alpha_\pi(t)}
A_{\gamma^*\pi^\to X}(M_X^2)\,,
\label{120}
 \eeq
where $M_X^2=(1-z)s$;  the 4-momentum
transfer squared $t$ has the form,
$t=-q_L^2- q_T^2/z$;
 and $\eta_\pi(t)$ is the phase (signature) factor which is nearly real in the vicinity of the pion pole.
The effective vertex function
$G_{\pi^+pn}(t)=g_{\pi^+pn}\exp(R_1^2t)$, where  
$g^2_{\pi^+pn}(t)/8\pi=13.85$. The value of the slope parameter $R_1$ is small \cite{kpp,kpps-dis} and is dropped-off for clarity in what follows.

Correspondingly, the fractional differential cross section of inclusive
neutron production in the Born approximation reads,
 \beq
\frac{1}{\sigma_{inc}}\,\frac{d\sigma^B_{p\to n}}{dz\,dq_T^2}=
\left(\frac{\alpha_\pi^\prime}{8}\right)^2
\frac{|t|}{z}\,g_{\pi^+pn}^2\left|\eta_\pi(t)\right|^2
(1-z)^{1-2\alpha_\pi(t)}\,
\frac{F_2^{\pi}(x_\pi,Q^2)}{F_2^p(x,Q^2)}\,,
\label{155n}
 \eeq
where $x_\pi=x/(1-z)$; $\alpha_\pi^\prime=0.9\GeV^{-2}$ is the pion Regge trajectory slope.

The Born approximation Eqs.~(\ref{120})-(\ref{155n}) is subject to strong absorption effects,
related to initial and final state interactions of the debris of the $\gamma^*\pi$ inelastic collision, which can be presented as two color-octet $\bar qq$ pairs, as is illustrated in Fig.~\ref{fig:pion-pole} (right). 
At high energies and large $z$ such a dipole should be treated as 
a 4-quark Fock component of the projectile photon,
$\gamma^*\to \{\bar qq\}_8-\{\bar qq\}_8$, which
interacts with the target proton via
$\pi^+$ exchange. This 4-quark state may also experience initial and final state
interaction via vacuum quantum number (Pomeron) exchange with the nucleons
(ladder-like strips in Fig.~\ref{fig:pion-pole} (right)). 

The absorption factor $S_{4q}(b)$ is naturally calculated in impact parameter representation,
relying on the well known parametrizations of the dipole cross section, measured at HERA.
The amplitude and the absorption factor factorize in impact parameters, then one should perform inverse Fourier transformation back to momentum representation. The details of this procedure can be found in \cite{kpps-dis}.

\begin{figure}[htb]
\centerline{
\includegraphics[width=4.5cm]{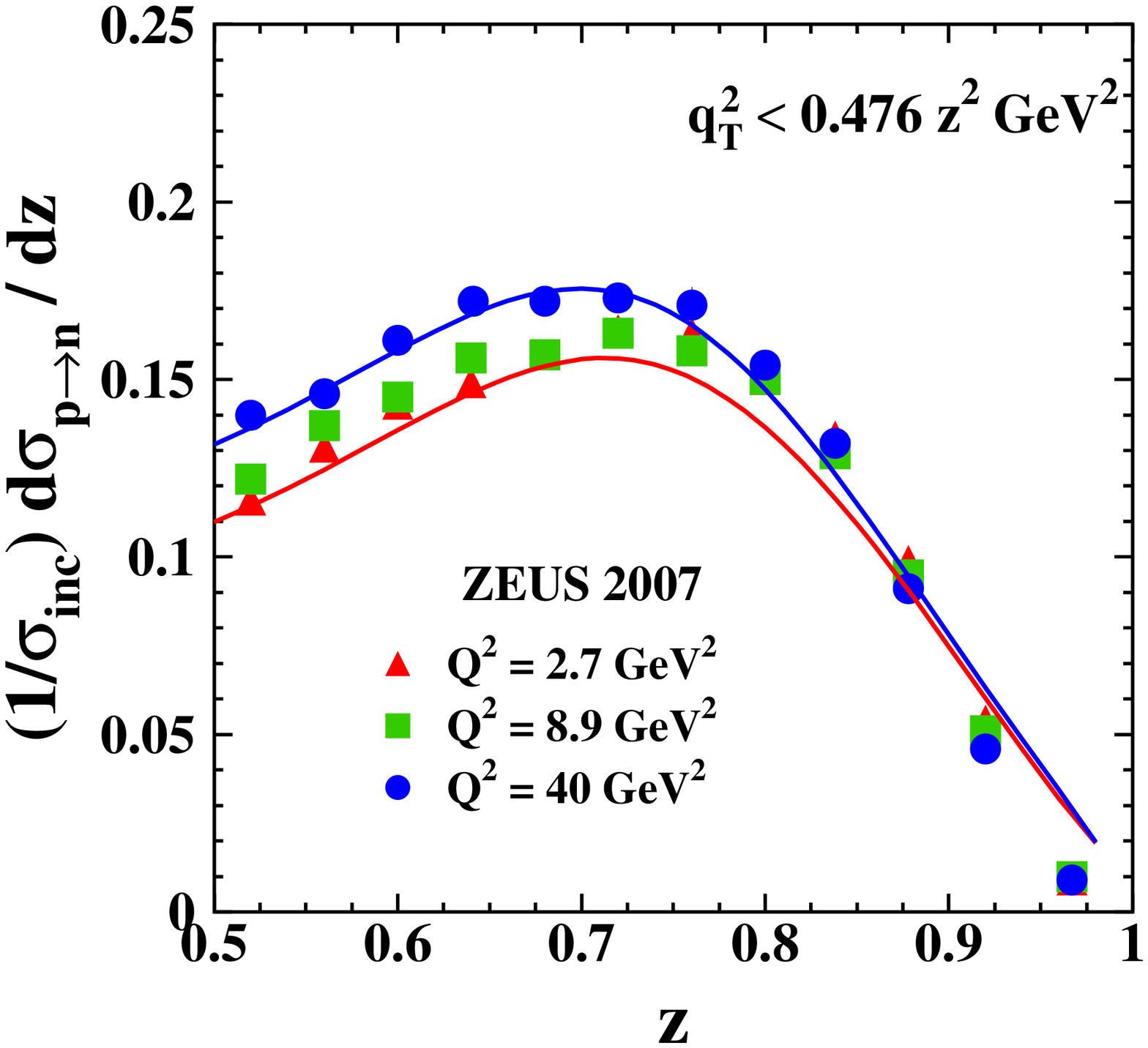}
\hspace{15mm}
\includegraphics[width=4.0cm]{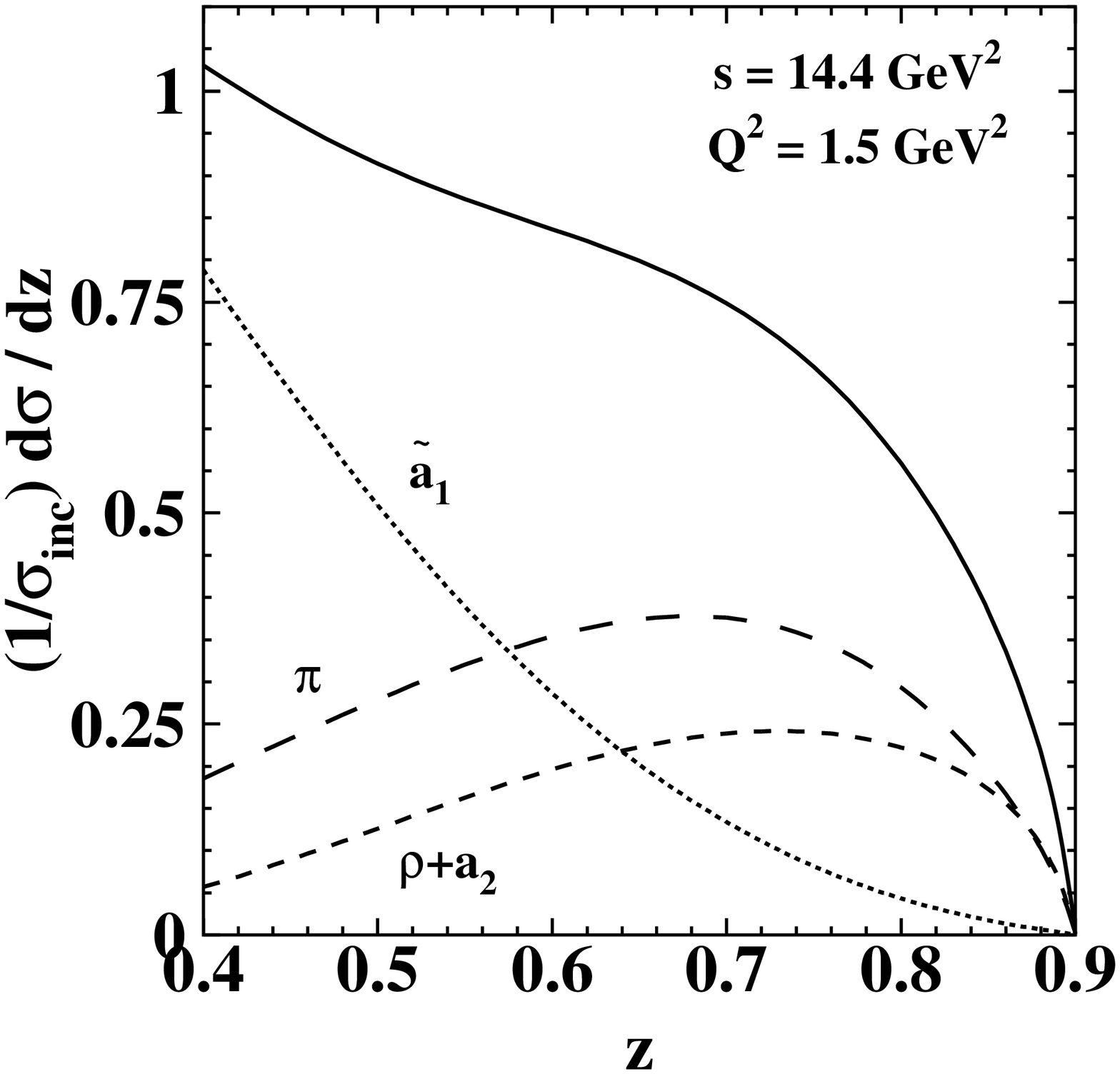}}
\caption{{\it Left:} comparison of the calculated $Q^2$-dependence of the fractional cross section of neutron production with data from \cite{zeus}. {\it Right:} the tractional cross section calculated at $Q^2=1.5\GeV^2$ and $\nu=8\GeV$.}
\label{fig:data-dis}
\end{figure}

The results of calculations are compared with data  \cite{zeus} on $Q^2$-dependence of the fractional cross section in Fig.~\ref{fig:data-dis} (left). 
%%%%%%%%%%%%%%%%%%%%
%%%%%%%%%%%%%%%%%%%%
The observed independence of $Q^2$ at large $z$ is a direct consequence of the mechanism of absorption under consideration, shown in Fig.~\ref{fig:pion-pole} (right). These results include, besides the pion exchange, also contributions from other iso-vector Reggeons, natural parity $\rho$ and $a_2$,
and unnatural parity $\tilde a_1$, which contains the weak $a_1$ pole and the strong $\rho$-$\pi$ Regge cut \cite{kpps-dis,kpss-AN}.

Analogous measurements of leading proton production from a neutron target (deuteron)
are planned to be done at Jefferson Lab. An example of expected fractional cross section 
calculated at $Q^2=1.5\GeV^2$ and $\nu=8\GeV$ is presented in Fig.~\ref{fig:data-dis} (right). The relative contribution of the pion pole is smaller compared with low-x processes, therefore the results are expected to be more model dependent.

\section{Single neutron production in $pp$ collisions}

\begin{figure}[htb]
%\vspace{1mm}
\centerline{
\includegraphics[width=4cm]{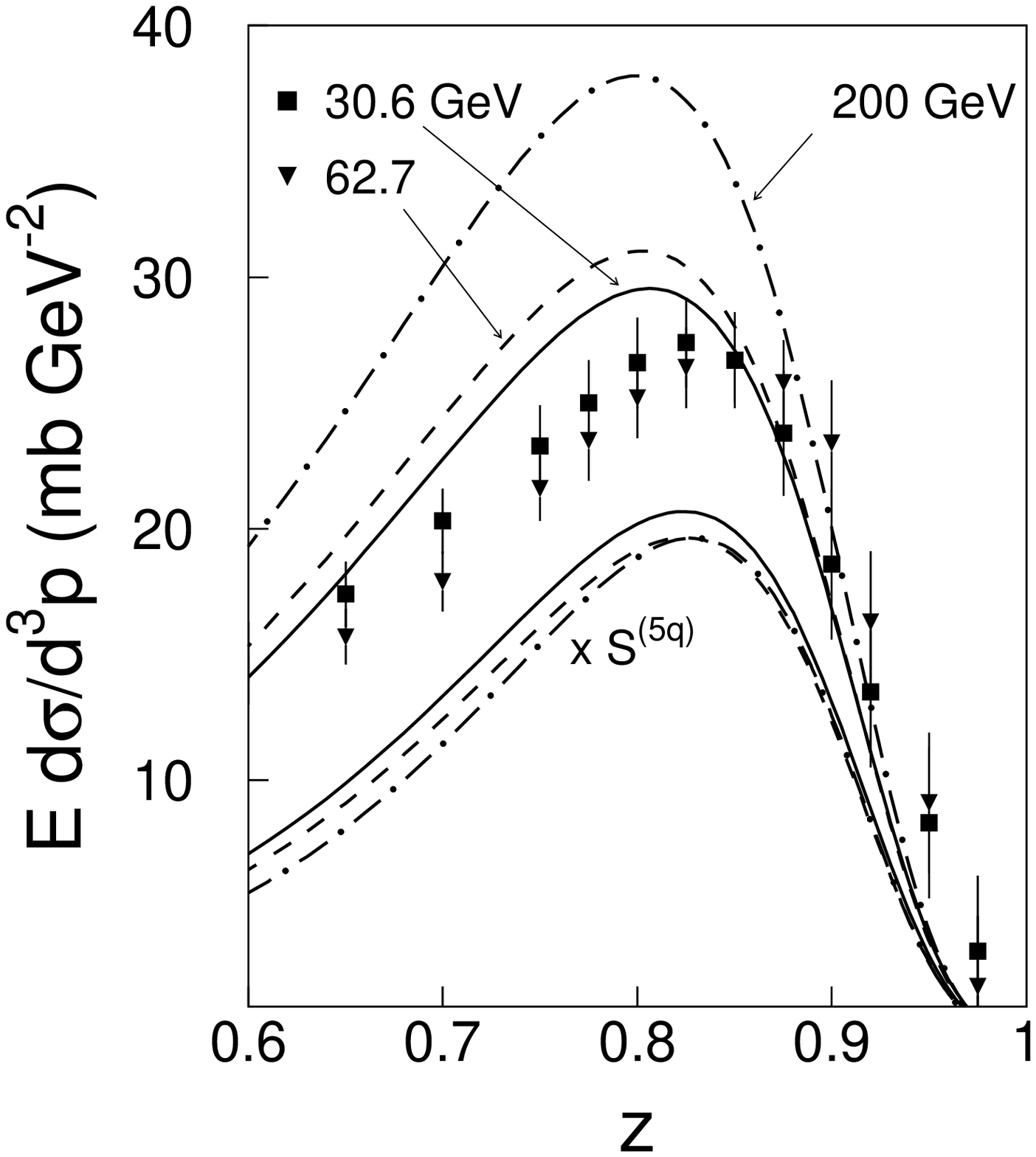}
\hspace{10mm}
\includegraphics[width=4cm]{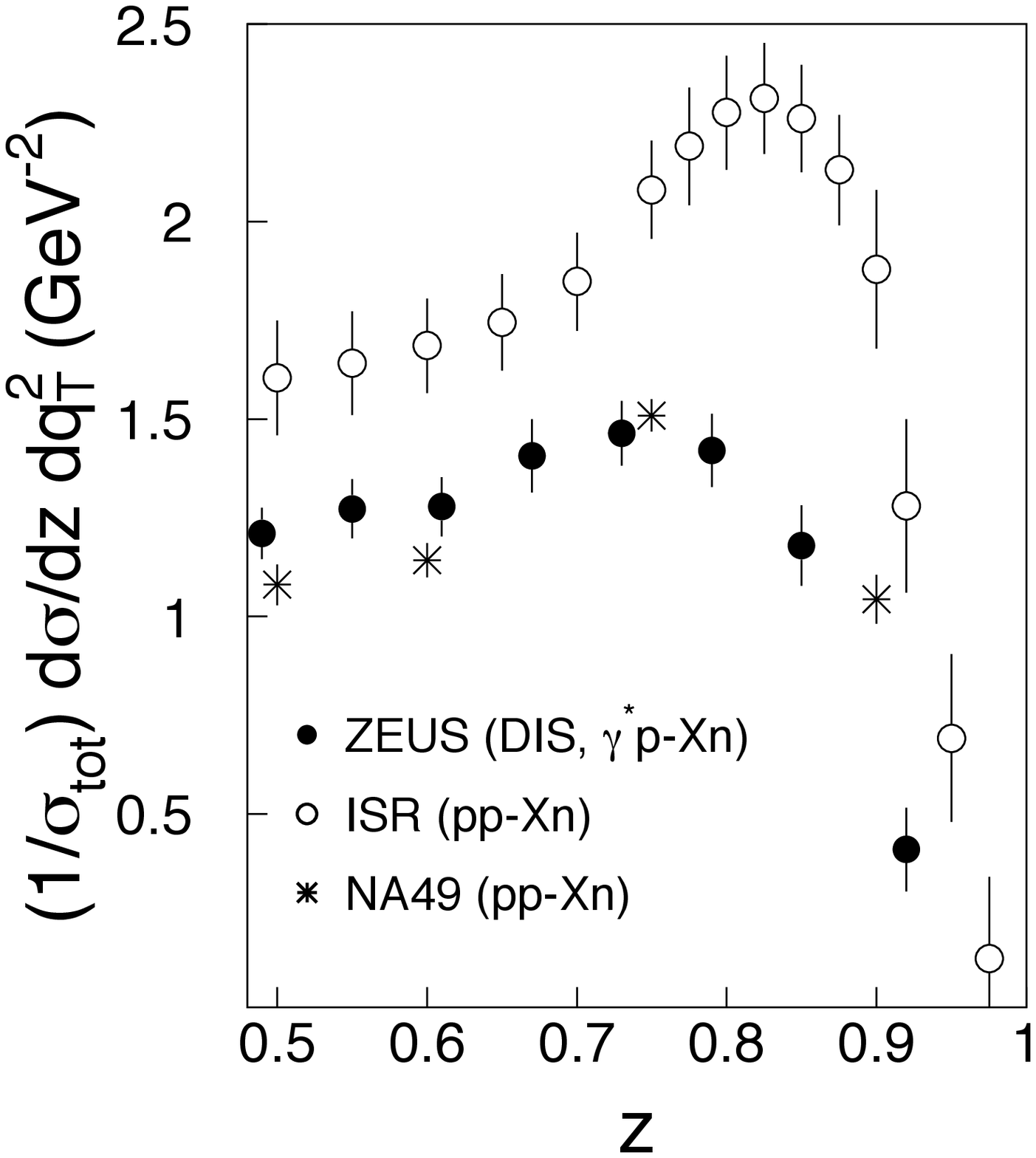}}
\caption{{\it Left:} energy dependence of the differential cross section of forward neutron production, calculated in the Born approximation (upper) and absorption corrected (bottom). Data are from \cite{zeus}. {\it Right:} Comparison of fractional forward cross sections of neutron production in $pp$ collisions \cite{isr,na49} and in DIS \cite{zeus}.}
\label{fig:data}
\end{figure}

Similar to DIS, production of leading neutrons at modern colliders (RHIC, LHC) offers a possibility to measure  
the pion-proton cross section at energies higher than has been available so far with pions beams. Otherwise, if the $\pi-p$ cross section is known or guessed, one can predict the cross section of $pp\to pX$, which is given by the same expression as Eq.~(\ref{155n}),
except the last factor, the ratio $F^\pi_2/F^p_2$, should be replaced by $\sigma^{\pi p}_{tot}(M_X^2)/\sigma^{pp}_{tot}(s)$. The results of  the Born approximation \cite{kpss} are depicted by upper three curves in Fig.~\ref{fig:data} (left), which  agree with ISR data at $\sqrt{s}=30.6$ and $62.7\GeV$ \cite{isr}. 
%%%%%%%%%%%%%%%%%%%%
%%%%%%%%%%%%%%%%%%%%
  
Of course these Born approximation results should be corrected for the absorption effects, which were found in \cite{3n,ryskin} rather weak, in agreement with the ISR data.
On the contrary, the absorption factor calculated in \cite{kpss} leads to a much stronger suppression, close to what was found for DIS in the previous section. As a consequence,
the absorption corrected three bottom curves in Fig.~\ref{fig:data} (left) strongly underestimate the data.

It was concluded in \cite{kpss} that the normalization of the data \cite{isr} is incorrect.
Indeed, comparison with other currently available data in DIS \cite{zeus} and in $pp$ collisions \cite{na49} plotted in Fig.~\ref{fig:data} (right), show that indeed these fractional cross sections are about twice below the ISR data. One hardly can imagine that absorption in photo-production process is stronger than in $pp$.

\begin{figure}[htb]
%\vspace{1mm}
\centerline{
\includegraphics[width=6cm]{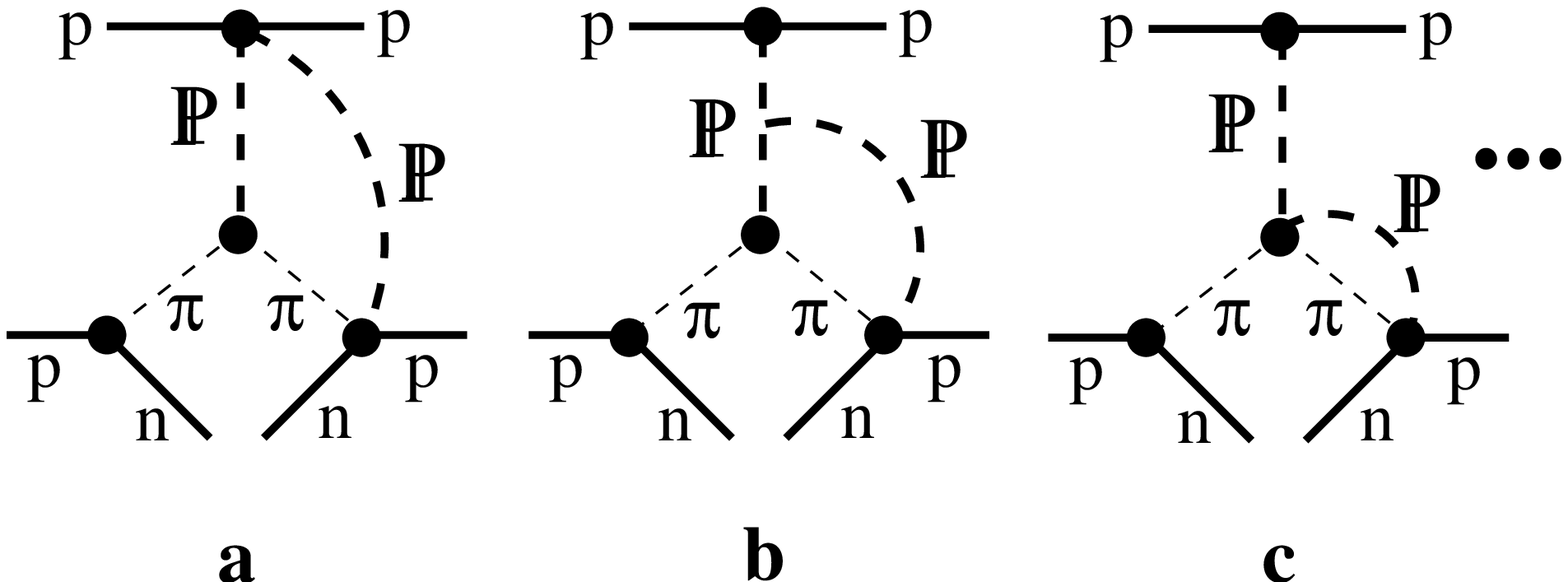}
\hspace{15mm}
\includegraphics[width=2.5cm]{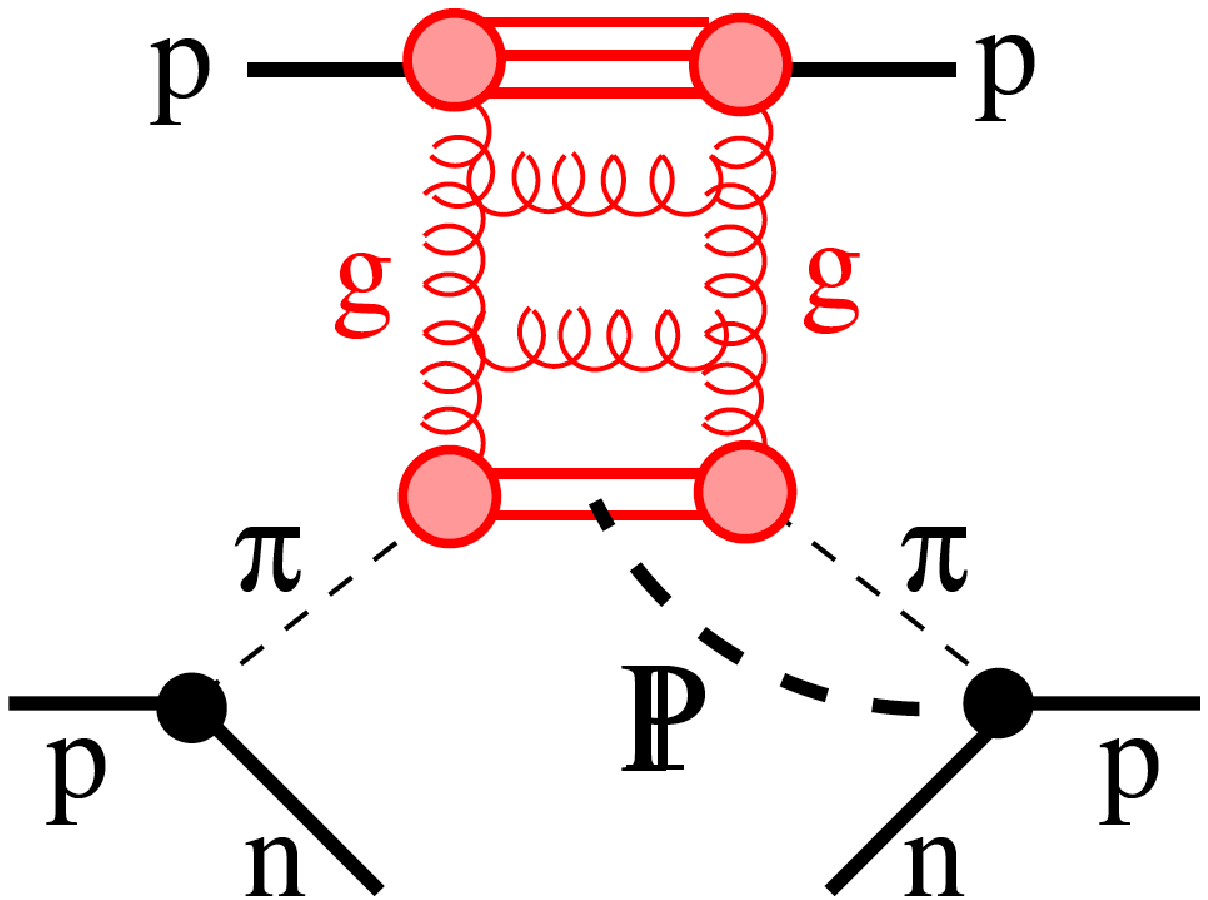}}
\caption{{\it Left:} triple-Regge graphs contributing to the absorption corrections. {\it Right:} the structure of the $2\Pom2\pi$ vertex in graph ({\bf c}).}
\label{fig:graphs}
\end{figure}

The reason of weak absorption found in \cite{3n,ryskin} is explained in Fig.~\ref{fig:graphs}. The third Reggeon graph ({\bf c}) was neglected because the 4-Reggeon vertex $2\Pom2\pi$
was claimed to be unknown. 
%%%%%%%%%%%%%%%%%%%%
%%%%%%%%%%%%%%%%%%%%
However,  this vertex has a structure, shown in the right part of Fig.~\ref{fig:graphs}, and it gives the largest contribution to the absorption effects.

In addition to the cross section, the rich spin structure of the amplitude, Eq.~(\ref{100}), suggests a possibility of a stringent test of the dynamics of the process, supported by recent precise measurements of the single-spin asymmetry of neutron production at RHIC \cite{phenix,phenix2}. Of course the amplitude (\ref{100}) does not produce any spin asymmetry because both terms have the same phase. However, the strong absorption corrections change the phase and spin effects appear. Nevertheless the magnitude of $A_N(t)$ was found in \cite{kpss-AN} to be too small in comparison with the data.

\begin{figure}[htb]
%\vspace{1mm}
\centerline{
\includegraphics[width=4.5cm]{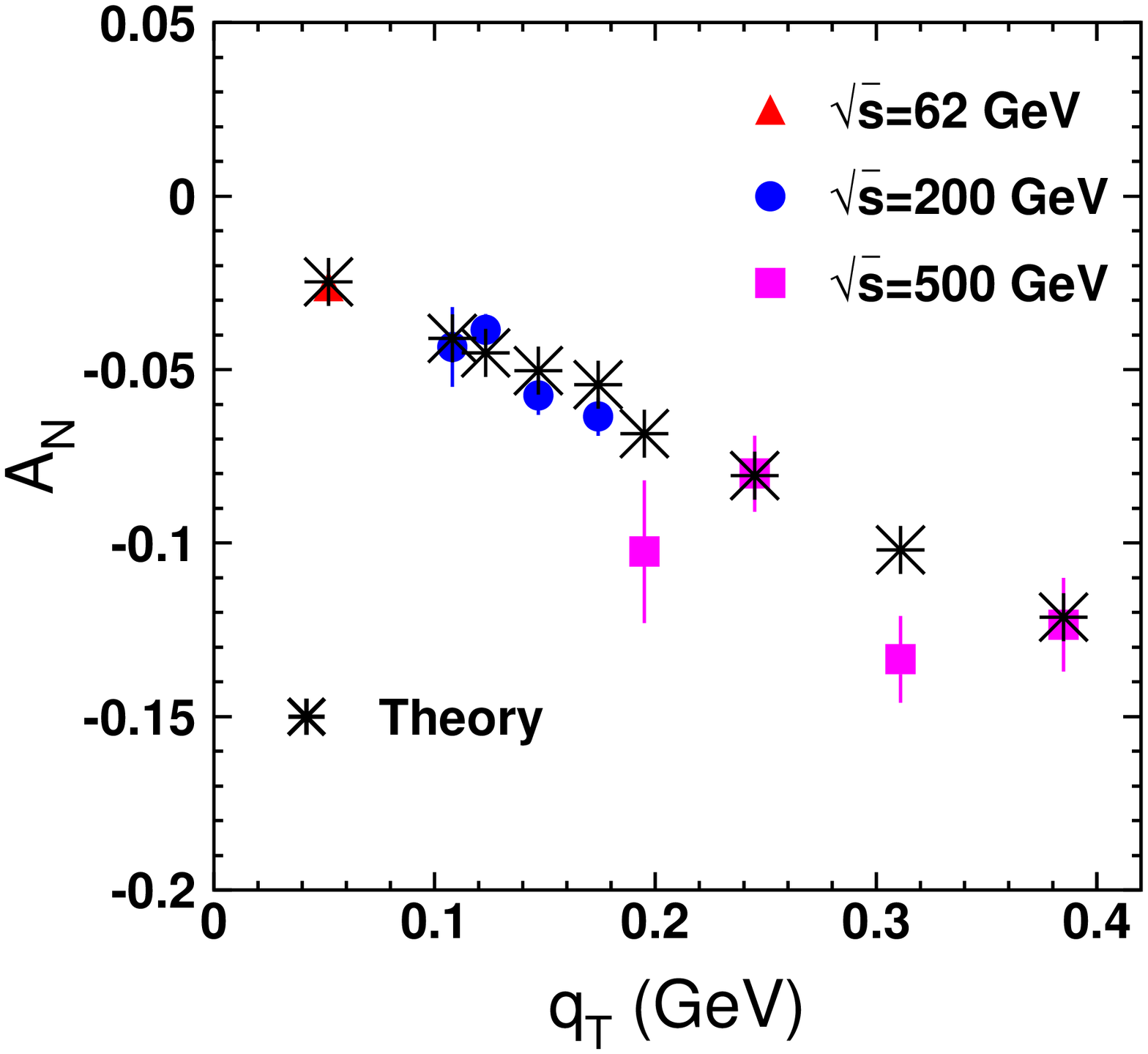}
\hspace{15mm}
\includegraphics[width=3.0cm]{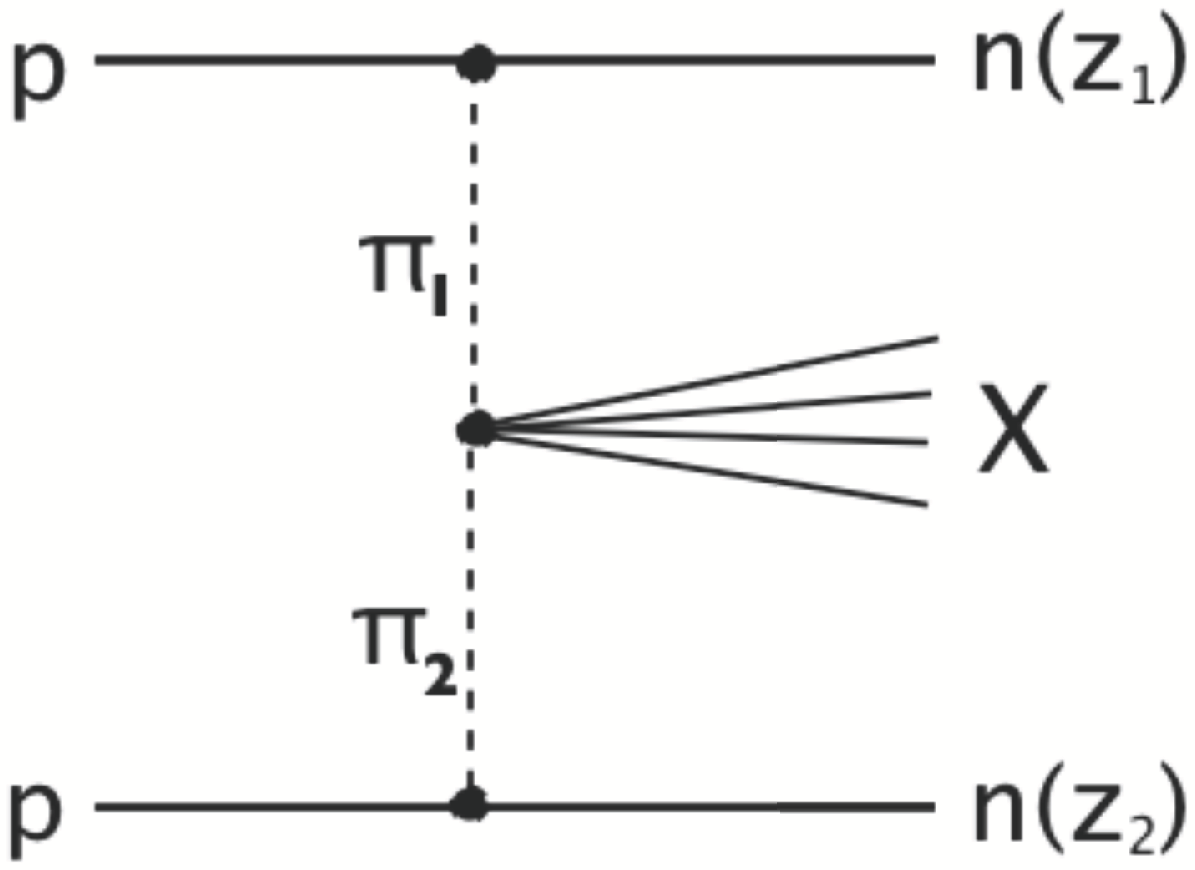}
}
\caption{{\it Left:} Comparison of the parameter-free calculations \cite{kpss-AN} (stars) with data \cite{phenix}. {\it Right:} double-pion exchange in the double-neutron production amplitude.}
\label{fig:AN}
\end{figure}

Interference with natural parity Reggeons $\rho$ and $a_2$ is strongly suppressed at high energies. Only the spin-non-flip axial-vector Reggeon $a_1$ is a promising candidate. Its effective contribution $\tilde a_1$ also includes the $\rho$-$\pi$ Regge cut. The results of parameter-free evaluation of $A_N(t)$ \cite{kpss-AN} due to $\pi$-$\tilde a_1$ interference, shown by stars in Fig.~\ref{fig:AN} (left),
well agree with data \cite{phenix,phenix2}.
%%%%%%%%%%%%%%%%%%%%
%%%%%%%%%%%%%%%%%%%%
\vspace{-5mm}

\section{Double-neutron production}

The large experiments at the RHIC and LHC colliders are equipped with zero-degree calorimeters (ZDC), which can detect small angle leading neutrons. This suggests an unique opportunity to detect two leading forward-backward neutrons. According to Fig.~\ref{fig:AN} (right)
one can extract from data  precious information about pion-pion interactions at high energies. 

\begin{figure}[htb]
%\vspace{1mm}
\centerline{
\includegraphics[width=4.8cm]{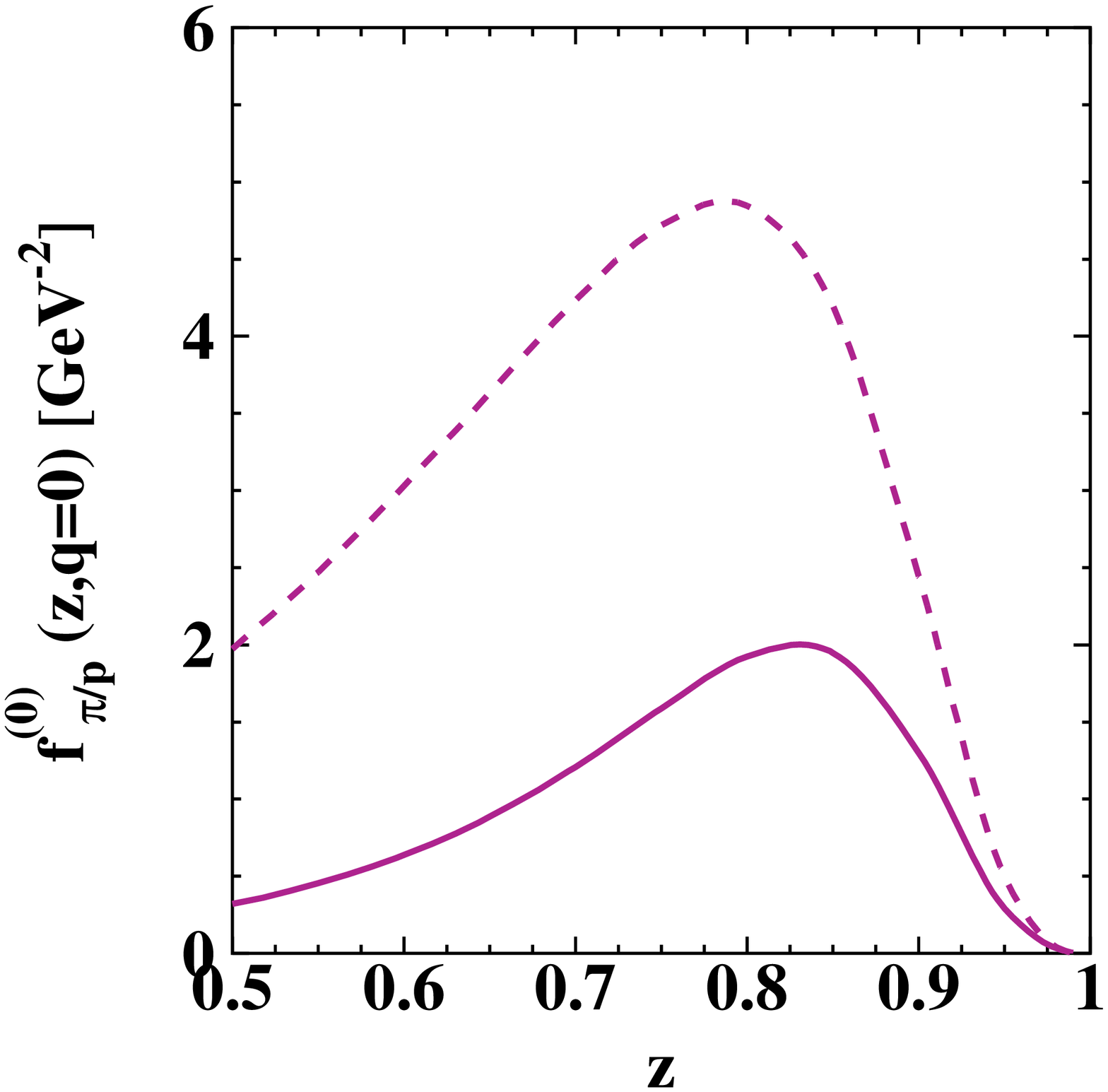}
\hspace{10mm}
\includegraphics[width=4.8cm]{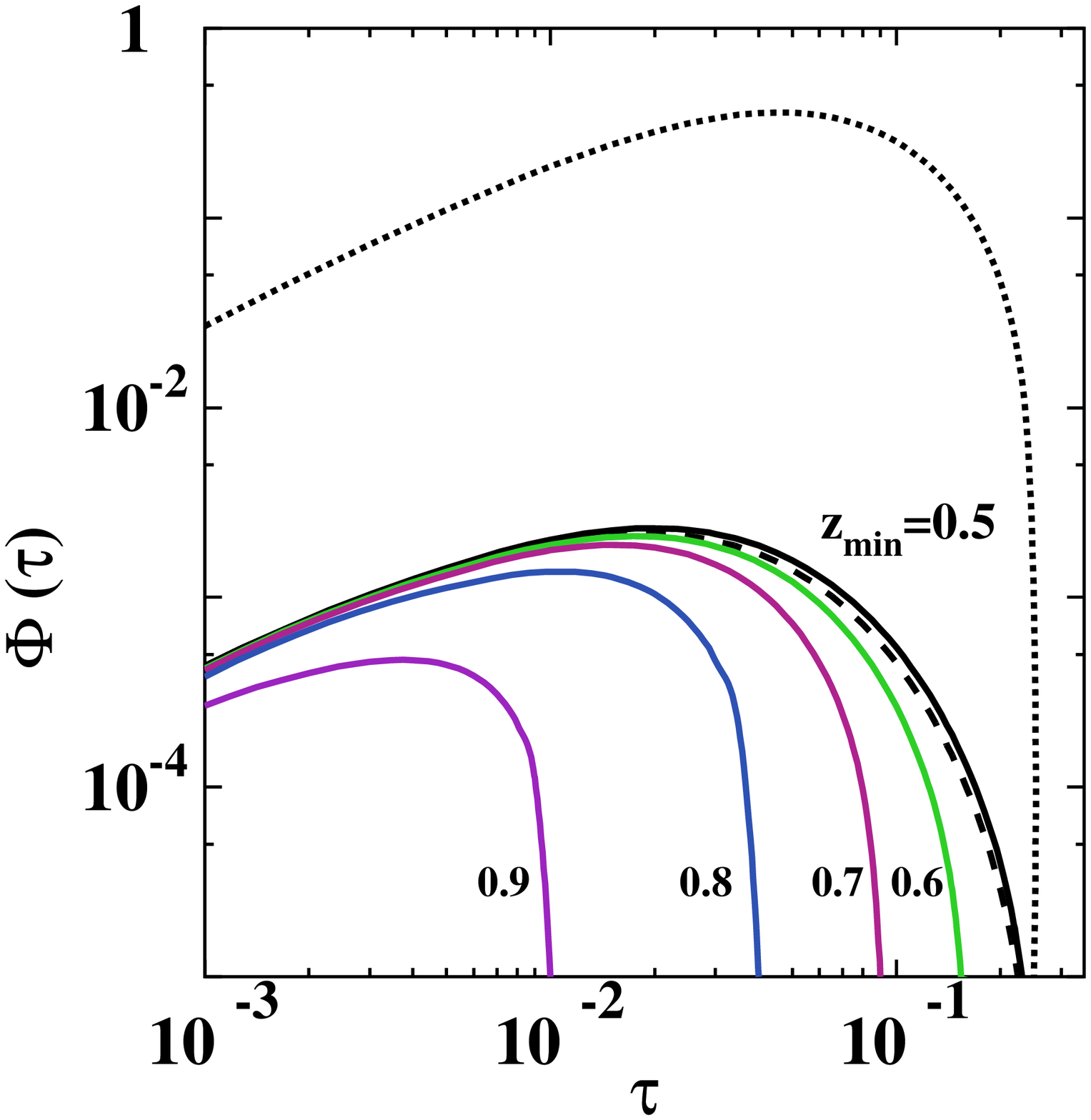}
}
\caption{{\it Left:} The forward flux of pions $f_{\pi^+/p}^{(0)}(z,q)$,
 calculated in the Born approximation (dashed) and including absorption (solid).
 {\it Right:} The integrated flux of two pions at $z_{min}=0.5$ in the Born approximation (dotted), and absorption corrected (dashed). Solid curves with $z_{min}=0.5$-$0.9$ also include the feed-down corrections.}
\label{fig:flux}
\end{figure}

The cross section of this process in the Born approximation has a factorized form \cite{pi-pi},
 \beq
\frac{d\sigma^B(pp\to nXn)}{dz_1dz_2\,dq_{1}^2dq_{2}^2}
= 
f^B_{\pi^+/p}(z_1,q_{1})\,
\sigma^{\pi^+\pi^+}_{tot}(\tau s)
f^B_{\pi^+/p}(z_2,q_{2}),
\label{160}
\eeq
where the pion flux in the proton with fractional momentum $1$-$z$ reads \cite{kpp},
\beq
f^B_{\pi^+/p}(z,q)=
-\frac{t}{z}\,G_{\pi^+pn}^2(t)
\left|\frac{\alpha_\pi^\prime\eta_\pi(t))}{8}\right|^2
(1-z)^{1-2\alpha_\pi(t)}.
\label{180}
 \eeq
This flux at $q=0$ is plotted by dashed curve in Fig.~\ref{fig:flux} (left).
%%%%%%%%%%%%%%%%%%%%
%%%%%%%%%%%%%%%%%%%%
The absorption corrected flux is plotted by solid curve, demonstrating a considerable reduction \cite{pi-pi}.

To maximize  statistics, one can make use of all detected neutrons, fixing $M_X^2=\tau s=(1-z_1)(1-z_2)s$
to extract the $\pi\pi$ total cross section, $\sigma(pp\to nXn)_{z_{1,2}>z_{min}}=
\Phi^B(\tau)\,
\sigma^{\pi^+\pi^+}_{tot}(\tau s)$. The double-pion flux $\Phi(\tau)$ reads,
\beqn
 \Phi(\tau)=
\frac{d\sigma(pp\to nXn)_{z>z_{min}}}
{\sigma^{\pi^+\pi^+}_{tot}(\tau s)}=
\int\!\!\! 
\frac{dz_1}{1-z_1}
F_{\pi^+/p}(z_1)
F_{\pi^+/p}(z_2)
D_{abs}^{NN}(s,z_1,z_2),
\nonumber
\eeqn
where $D_{abs}^{NN}(s,z_1,z_2)$ is an extra absorption factor due to direct $NN$ interactions, which breaks down the pion-flux factorization. This factor was calculated in \cite{pi-pi}. The integrated pion flux $F_{\pi^+/p}(z)$ reads,
\beq
F_{\pi^+/p}(z)=-z\int\limits_{q_L^2}^\infty dt\,
 f_{\pi^+/p}(z,q).
 \label{197}
 \eeq 

Thus, detecting pairs of forward-backward neutrons with the ZDCs installed in all large experiments at RHIC and LHC, provides an unique opportunity to study the pion-pion interactions at  high energies. However, the absorption effects are especially strong for this channel.

\section{Summary}

The higher Fock components of the proton, containing pions, allow to get unique information about the pion structure and interactions at high energies, provided that the kinematics of neutron production is in the vicinity of the pion pole. In this short overview we presented several processes with neutron production, DIS on a proton, proton-proton collisions, including spin effects, and also double neutron production. However, even in a close proximity of the pion pole,
the analysis  can hardly be performed in a model independent way. Strong absorption effects 
significantly suppress the cross sections. We identified the main mechanism for these effects, which has been missed in previous calculations. It arises from initial/final state interaction
of the debris of the pion collision. It was evaluated employing the well developed color-dipole phenomenology, based on DIS data from HERA.\\

{Acknowledgments:} B.Z.K. is thankful to the organizers of the EDS Blois Meeting
for inviting to speak.
This work was supported in part
by Fondecyt (Chile) grants 1130543, 1130549 and 1140390.

\end{document}